# An Investigation on the Cyclic Temperature-Dependent Performance Behaviors of Ultrabright Air-Stable QLEDs

*Saeedeh Mokarian Zanjani, Sadra Sadeghi, Afshin Shahalizad, Majid Pahlevani*[*]

Dr. S. Mokarian Zanjani, Dr. S. Sadeghi, Prof. M. Pahlevani*

Department of Electrical and Computer Engineering, Queen's University, Kingston, Ontario, K7L 3N6, Canada

Dr. A. Shahalizad*

Genoptic LED Inc., Calgary, Alberta, T2C 5C3, Canada

[*]E-mail : majid.pahlevani@queensu.ca

**Abstract**

The aerobic and thermal stability of quantum-dot light-emitting diodes (QLEDs) is an important factor for the practical applications of these devices under harsh environmental conditions. In this paper, we demonstrate all-solution-processed amber QLEDs with an external quantum efficiency (EQE) of >14% with almost negligible efficiency roll-off (droop) and a peak brightness of >600,000 cd/m$^2$, unprecedented for QLEDs fabricated under ambient air conditions. We investigate the device efficiency and brightness level at a temperature range between -10℃ to 85℃ in a 5-step cooling/heating cycle. Unlike previous studies reported in the literature, we conducted the experiments at relatively high brightness levels, required for outdoor lighting applications. The results reveal that the device performance increases slightly at sub-zero temperatures (-10℃) and drops slightly at very high temperatures (85 ℃), proving acceptable thermal stability. Overall, the performance parameters do not change dramatically over the temperature range within the experimental uncertainty range. Interestingly, the device efficiency parameters recover to the initial values upon returning to room temperature. The variations in the performance are correlated with the modification of charge transport characteristics and induced radiative/non-radiative exciton relaxation dynamics at different temperatures. Being complementary to previous studies on the subject, the present work is expected to shed light on the potential feasibility of realizing aerobic-stable ultrabright droop-free QLEDs and encourage further research for solid-state lighting applications.

## 1. Introduction

Quantum-dot light-emitting diodes (QLEDs) has sparked a significant amount of attention from both academia and industries, owing to their exceptional optoelectronic properties, making them suitable for various electronic devices. [1-7] For example, external quantum efficiency (EQE), [8-11] brightness level, [12,13] and operational lifetime [14-17] of QLEDs have now reached the standards for commercial QD-based displays.[18] On the other hand, taking advantage of luminescent colloidal quantum-dots (QDs) with well-engineered graded multi-shell configurations, negligible EQE roll-off (droop) of QLEDs at high brightness levels[19] may allow for their applications in outdoor lighting, projection display, and phototherapy. [19-21] However, due to issues such as operational stability, shelf stability, and efficiency roll-off, reliable high-brightness QLEDs suitable for solid-state lighting systems are still far away from commercialization.[21]

In that context, thermal stability at high brightness levels with preserved efficiency under harsh environmental conditions (i.e., very high/low temperatures and high humidity levels) is a key factor for outdoor LED lighting systems. Device protection against oxygen and moisture at high humidity levels (e.g., 85 RH) can be ensured by employing advanced thin-film encapsulation methods,[22] but the thermal stability of an LED system depends on both intrinsic and modified charge transport characteristics in the functional charge transport layers[23] as well as exciton relaxation dynamics in the emissive layer (EML). Several studies have been published previously on the temperature-dependent electroluminescence (EL) performance of organic LEDs (OLEDs),[24-26] perovskite LEDs,[27,28] and QLEDs[29-31]. As the focus of the present work, for example, M. Zhang et al. investigated both photoluminescence (PL) and EL performance of their red QLEDs within a 120-300 K temperature range but did not carry out their experiments at temperatures higher than room temperature (RT).[31] The authors reported enhancement of the current density and reduction of turn-on voltage as the temperature was increased to RT from sub-zero temperatures. In a study by J. Yun et al., the authors explored the current density vs. voltage (J-V) behavior of their inverted colloidal Cd-based QLEDs at 100-400 K but did not report efficiency parameters.[29] Biswas et al. implemented a heat-assisted method for improving the EQE and current efficiency (LE) of their colloidal CuInS-based yellow QLEDs.[30] These authors observed that by increasing the substrate temperature during the sputtering of ZnO ETL, the charge injection of their devices improved, which led to an efficiency enhancement. Recently, Sue et al. proposed that the root cause of up-conversion EL (i.e., sub-bandgap turn-on EL) typically observed in QLEDs is due to thermally-assisted charge injection by exposing their devices to a broad temperature range. However, these authors performed their experiments within a small voltage range (around the turn-on voltage) and apparently did not investigate the temperature-dependency at high brightness levels.[32] Additionally, one should note that, even without exposure to external high/low temperatures, the operating temperature of QLEDs at high brightness levels typically exceeds RT, due to the high current passing through the device, causing Joule heating.[33-35] Therefore, proper thermal management is crucial to long-term operational stability of high-brightness QLEDs.

Herein, we report on solution-processed air-stable amber QLEDs (EL peak at 600 nm) with extremely narrow emission linewidth and study their thermally-induced optoelectronic behaviors. Our devices show a high maximum EQE of over 14% and a maximum brightness of 624000 cd/m$^2$ at 12 V, unprecedented for QLEDs fabricated under ambient-air conditions (see, e.g., Ref. [36]) [37]. We have experimentally investigated the temperature-dependent efficiency parameters and brightness by stressing the devices under multiple cooling/heating cycles within a wide temperature range from -10℃ to 85℃. Unlike some results previously published in the literature, [29]no device failure was observed after heating the devices up to 85℃. Specifically, we operated our devices in a 5-step thermal cycle by starting from RT1, cooling down to -10℃, returning to RT2, heating up to 85℃, and finally returning to RT3, sequentially. Compared to RT, results reveal a slight increase in the efficiency at -10℃ and a slight reduction of efficiency at 85 ℃ (Figure 1). However, the overall device efficiency remains almost unchanged within the experimental range of uncertainty. The biggest change is observed in terms of the brightness level, which is significantly affected by the applied temperature. The charge transport characteristics were also modified at different temperatures, increasing and reducing the turn-on voltage at low and high temperatures, respectively. Furthermore, the PL spectra of the QDs thin films were recorded at different temperatures to be able to distinguish between optical and electrical effects. Particularly, a temperature-dependent spectral shift in the emission peak and intensity was distinctly observed in both PL and EL. The present work provides insights into the feasibility of realizing air-stable, thermally-stable, and efficient QLEDs potentially suitable for practical outdoor lighting and display applications.

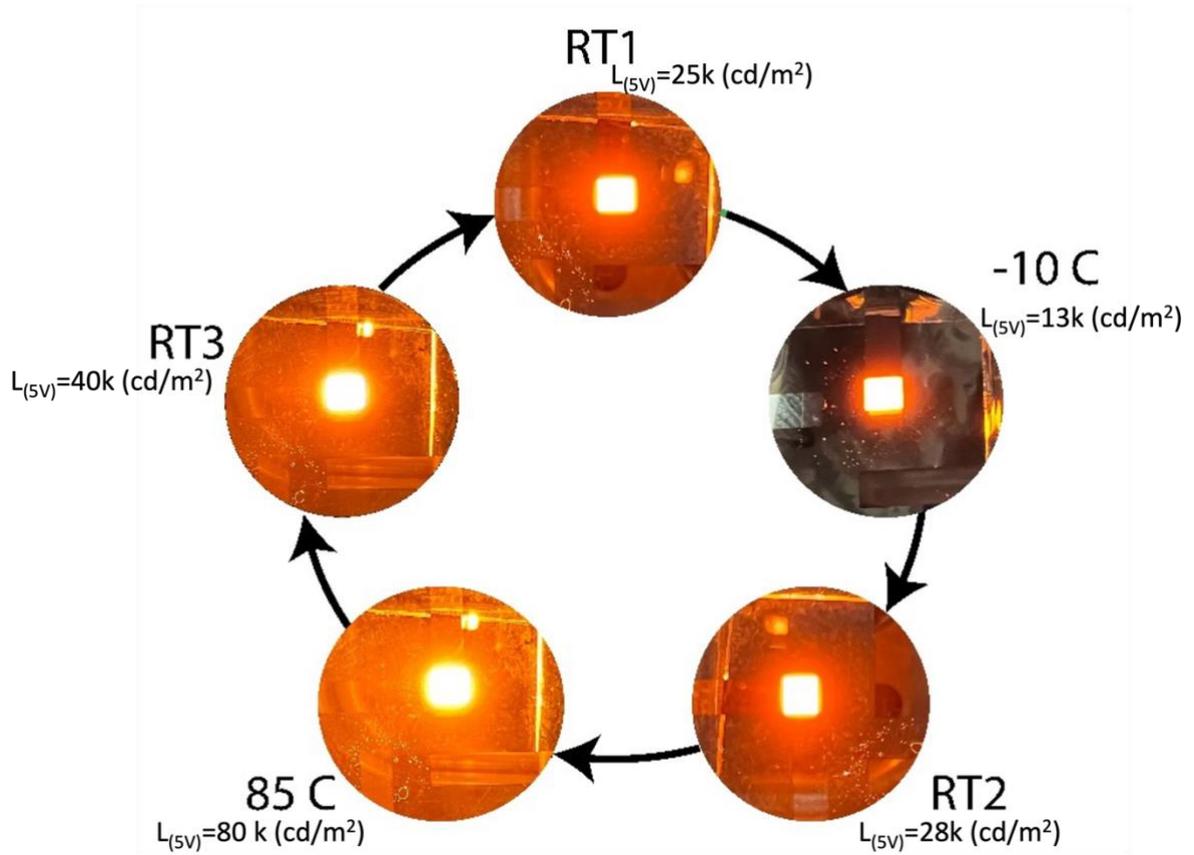

**Figure 1.** Photographs of an amber bottom-emitting QLED operated under heating/cooling cycles.

## 2. Materials and QLED device characterization

We synthesized ZnCdSe/ZnSe/ZnSeS/ZnS Amber quantum dots (QDs) as the emissive layer (EML) and 10%Li:10%Mg:ZnO (LMZO) nanoparticles (NPs) as the electron transport layer (ETL) in the QLED structure. Since the focus of the present work is primarily on temperature effects, details of the QDs synthesis with full characterizations will be published elsewhere. The characterizations of QDs and LMZO are demonstrated in Figure 2a-c.

Figure 2a shows the PL and absorption spectrum of the QDs, where the PL peak is centered at 599 nm with the full width at half maximum (FWHM) of 21 nm. The QDs solution photoluminescence quantum yield (PLQY) measurements showed an average of 97% ± 2, attributed to the strong confinement of electrons and holes inside the ZnCdSe core, which led to the efficient core emission instead of the emission from the other shell materials. Additionally, the formation of the ultra-thick ZnSe/ZnSeS/ZnS shell (14 monolayers) on the core QDs effectively suppressed the Förster resonance energy transfer (FRET), leading to a high film PLQY value of 70%. The absorption spectrum obtained from UV-Vis spectroscopy demonstrates the first excitonic peak position at 597 nm for the QDs (blue curve) and 294 nm for LMZO NPs (pink curve) in Figure 2a and 2b, respectively. The LMZO bandgap is calculated to be 3.9 eV, based on the method used in reference.[38] The schematic of the spin-coated QLED device, energy band alignment, and the cross-sectional TEM image of the layer stack are shown in Figure 2c-2e.

The layers are spin coated on patterned ITO/glass substrates with a sheet resistance of 15-20 Ohm/sq. Poly(ethylene dioxythiophene): polystyrene sulfonate (PEDOT: PSS) with the thickness of 80 nm was used as the hole injection layer (HIL). 40 nm-thick Poly[(9,9-dioctylfluorenyl-2,7-diyl)-co-(4,4'-(N-(p-butylphenyl)) diphenylamine)] (TFB) was used as the hole transport layer (HTL). The QD layer with the thickness of 35 nm, and the 110 nm-thick LMZO were respectively the EML and ETL in the device structure. Finally, 80 nm of aluminum (Al) and 175 nm of silver (Ag) were deposited by thermal evaporation to serve as cathode. A UV-curable epoxy resin together with glass coverslips were used for the encapsulation before characterizing the devices in air.

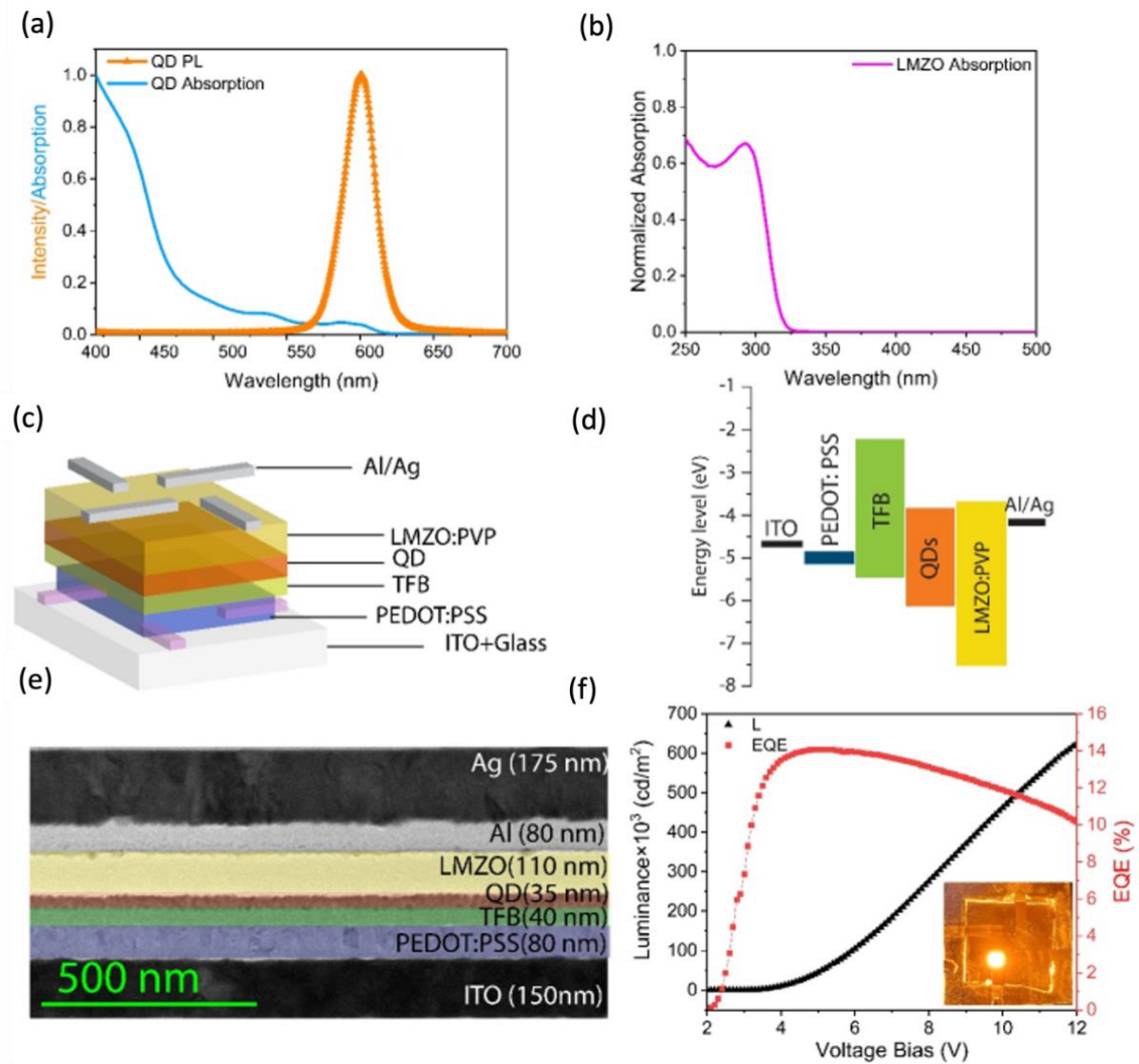

**Figure 2.** (**a**) Absorption and PL spectra of the QDs, (**b**) absorption spectrum of LMZO in solution (**c**) layer-stack of the QLED (**d**), energy level alignments of the layer-stack (**e**), cross-sectional TEM image of the QLED structure (**f**), and performance parameters. The inset in (g) shows the photograph of a QLED device operated at 12 V.

## 3. Results and Discussions

### 3.1. Temperature-dependent PL and EL characteristics

In this section, first, we discuss the effect of cooling/heating cycles on the PL and EL characteristics of the QDs film and QLEDs, respectively. Figure 3a shows the PL spectra of the QDs films (excited at 365 nm) at different temperatures. According to Figure 3a and 3c, the PL peak position undergoes a blue-shift at -10℃. Specifically, the PL peak shifts from 601 nm at RT1 to 598 nm and the spectral linewidth becomes narrower at -10℃ (from 26.5 at RT1 nm to 25.4 nm), and then it retrieves to 601 nm when the film equilibrates to RT2. The origin of such blue-shift in the PL peak has been previously found to be bandgap expansion of QDs at low temperatures. [31,39-41] An opposite effect is also observed when the temperature rises to 85 ℃, where the PL peak position undergoes a red-shift to 606 nm and the linewidth broadens (from 26.4 nm at RT2 to 29.3 nm), which are respectively ascribed to bandgap shrinkage and enhanced carrier-phonon scattering in QDs nanocrystals (i.e., the ZnCdSe core in our case) and subsequent defect emission at elevated temperatures. [42-45] As observed after the third cooling cycle, the peak position reversibly returns to 601 nm as the film equilibrates back to RT3, but the linewidth (27 nm) remains slightly wider than that at RT2.

On the other hand, the PL intensity enhances at -10 ℃, which can be attributed to the suppression of the non-radiative recombination channels through partial elimination of the temporary trap states formed due to the ion-exchange process. [34] Additionally, the probability of carrier localization for radiative recombination is higher at lower temperatures. In contrast, the carrier trapping (and subsequent non-radiative recombination) is expected to enhance at elevated temperatures due to formation of permanent and temporary trap states. This is clearly seen in the declined PL intensity at 85 ℃. The partial recovery of the PL intensity upon cooling back to RT3 indicates relaxation of temporary trap states. The irreversible portion of the PL intensity can also be explained by emission quenching due to the formation of permanent trap states at elevated temperatures. [42] It is known that detachment of caping ligands in QDs solutions at elevated temperatures can contribute to formation of surface trap states.[46] However, in our case, since the 1-Octanethiol (OT) surface ligands are strongly bound to the QDs surface and considering that the measurements were done with solid-state films, it is unlikely that OT ligand detachment could cause significant exciton-quenching due to surface traps. It would not also be expected that the elevated temperature (85 ℃) in our experiments can be sufficient for detaching the capping ligands because studies have shown that an onset temperature of around 100 ℃ would be needed.[34] Therefore, it may be expected that such temperature-induced trap states, which are casing the irreversible quenching, form in the shells or at the interfaces in the QD composition.

Figures 3b and 3d show the temperature-dependent EL spectral behaviors of an amber QLED device recorded at a fixed driving voltage. A similar spectral shift trend was also observed in the EL behavior in the cooling/heating cycles under the same experimental conditions. However, the changes in the emission intensity and spectral linewidth were found to be distinctly different. Specifically, the EL spectrum measured initially at RT1 (603 nm) showed a small 2 nm red-shift compared to the PL peak, but the linewidth remained the same. A blue-shift to 600 nm with a linewidth of 25.4 nm (similar to the PL) was observed when the device was cooled down to -10 ℃ and then retrieved when equilibrating back to RT2. Zhang et al. ascribed such a spectral blue-shift and linewidth narrowing to carrier relaxation into lower energy localized states, and the change in charge-recombination dynamics at reduced temperatures.[31] The EL peak also red-shifted to 606 nm and a much broader linewidth of 37.4 nm was observed compared to the PL at 85 ℃. Contrary to the PL behavior, however, the EL intensity declined at -10 ℃ and was boosted significantly at 85 ℃. After heating and cooling back to RT3, similar to the PL behavior, the EL intensity retrieved partially but not fully to their initial levels, likely due to the trap states discussed earlier.

The opposite trends observed in the PL and EL behaviors reflects that temperature-dependent charge injection (into the QDs) and charge transport modifications have a more impact than the effect of temperature at the QD level. For instance, it has been previously reported that the trapped electrons at elevated temperatures are released from the trap states and the charge transport and radiative recombination are enhanced.[31] To elaborate the observed effects, as discussed in the following, we investigated the variations in the temperature-dependent efficiency parameters, brightness level, and current density-voltage (J-V) behaviors in our QLEDs (Figure 4 and Table 1).

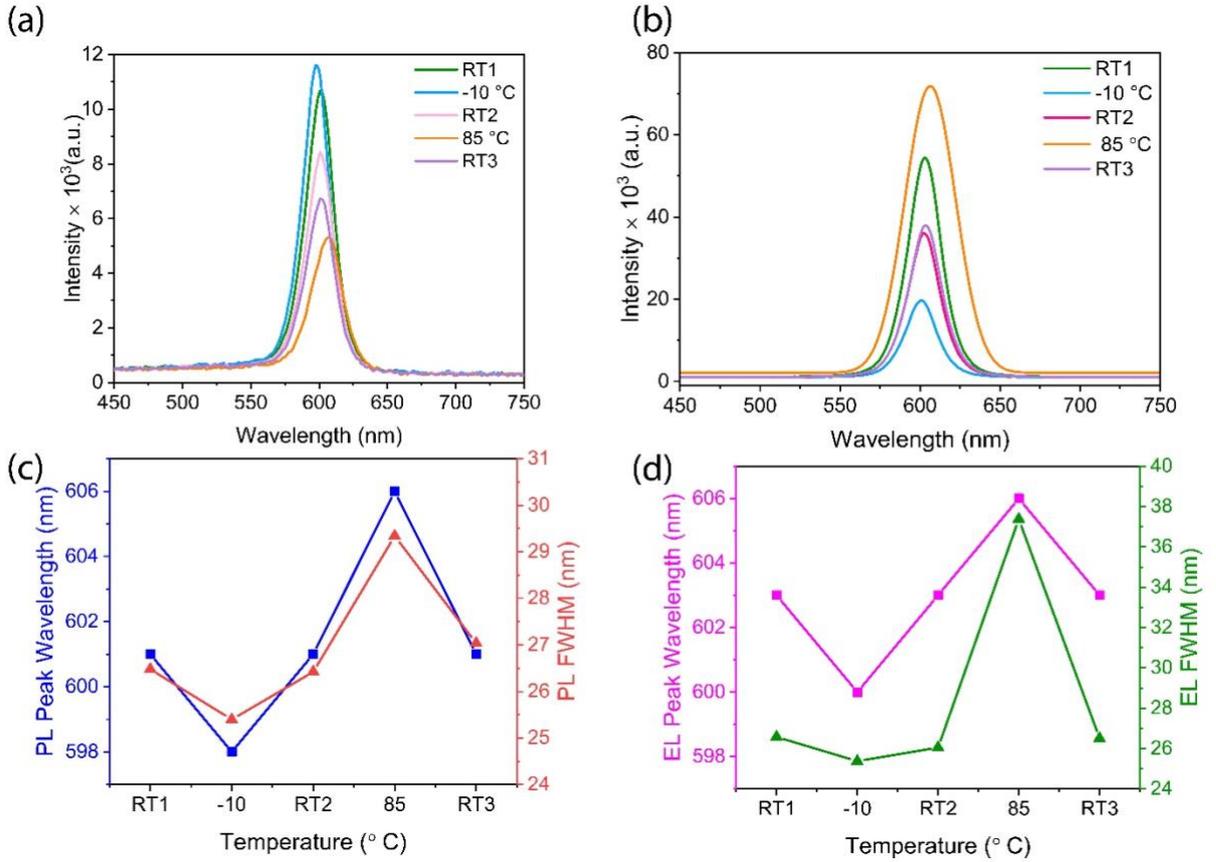

**Figure 3.** Temperature-dependence of **(a)** PL of the QDs film, **(b)** EL of the QLEDs recorded at a fixed driving voltage of 5V, **(c)** PL and **(d)** EL peak full width at half maximum (FWHM) and spectral position at -10℃ to 85℃.

The initial brightness level at RT1 was recorded to be >20,000 cd/cm$^2$ at 5 V. Temperature-dependent EL studies have not been reported at such a high brightness level in literature. Figure 4a shows the J-V plots of the QLED at various temperatures. When the device was first cooled down to -10°C from RT1, the current density declined, but when it was equilibrated back to RT2, the current density retrieved almost to its initial value at RT1. In contrast, the current density values of the device showed that the charge transport properties were reinforced at the elevated temperature of 85°C. Owing to thermally-assisted charge injection at elevated temperatures, [28] the turn-on voltage ($V_{on}$) was reduced from 2.1 V at RT to 1.8 V at 85℃. The $V_{on}$ was also higher (2.3 V) at -10°C, due to a reverse effect. In addition to thermally-assisted charge injection, given that the efficiency parameters drop slightly and considering that the brightness increases substantially at 85°C (Figure 4b), the dramatically increased current density may also be correlated with increased leakage current within the voltage range in our experiments. Specifically, even though the current density at RT3 was lower than that at 85°C, it was still higher compared to its initial value at RT1.

Given that the efficiency parameters do not completely return to the values at RT2, this may be due to any plausible physical damage (due to the leakage current) occurring to the polymer hole transport layer (HTL) in the device structure. Such a physical damage at the HTL/QDs interface has been reported to be one of the main reasons for QLED degradations at high brightness/current density levels.[47] Nevertheless, in our case, the likely damage to the HTL does not seem to be severe because the efficiency parameters return almost to their initial values at RT after the cooling/heating cycles, indicating thermal stability of the devices withing the temperature range. Table 1 summarizes the QLED performance parameters operated in a thermal cycle.

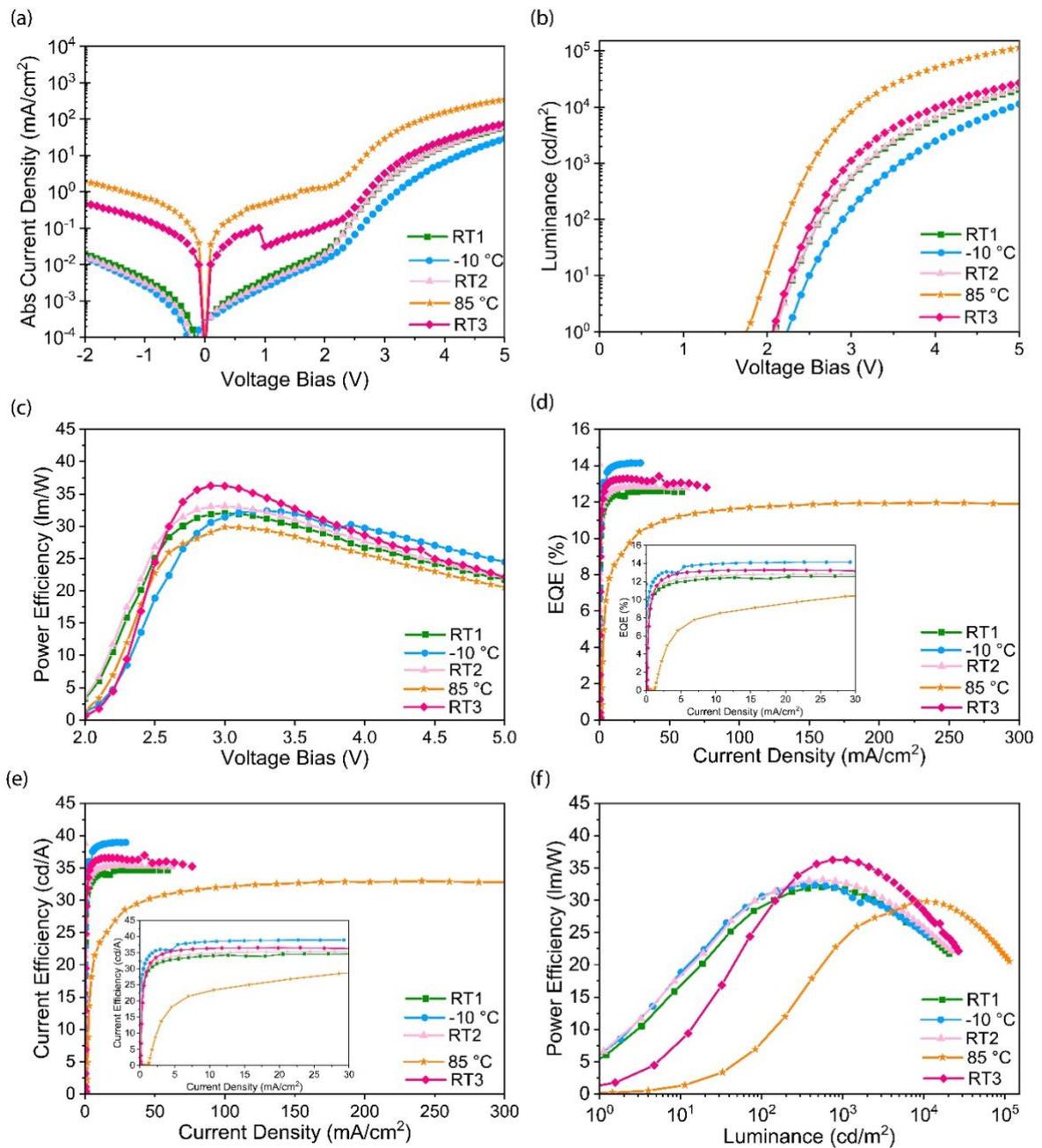

**Figure 4.** Amber QLED device performance at reduced and elevated temperature cycles. **(a)** Current density (J) **(b)** Brightness (L) and **(c)** Power efficiency (PE) versus voltage. **(d)** EQE$_{max}$ vs. current density (J) **(e)** Current efficiency (CE) vs. current density (J) **(f)** Power efficiency (PE) versus brightness (L). The insets of Figure 3d and 3e show the magnified x- axis to better demonstrate the current density values at T<85℃.

The maximum power efficiency (PE$_{max}$) shown in Figure 4c and 4f were also found to be the lower at 85℃ compared to those recorded at RT, due to the increased turn-on voltage. On the other hand, the maximum current efficiency (LE$_{max}$) was higher for the QLED operating at -10℃ (Figure 4e), which is in direct contrast with the previously reported results from Zhang et al. [31] In addition, the EQE$_{max}$ value was found to be higher at -10℃ (Figure 4d), which may be explained by the expected suppressed non-radiative recombination channels at reduced temperatures. One should note that the different lengths of the curves in Figure 4d, and 4e at various temperatures originate from the variations in the maximum current density values at 5 V (x-axis).

**Table 1.** QLED device performance parameters at different temperatures. The voltage was swept up to 5 V.

| T [℃] | RT1 | -10 ℃ | RT2 | 85 | RT3 |
|---|---|---|---|---|---|
| **V$_{on}$ [V]** | 2.1 | 2.3 | 2.1 | 1.8 | 2.1 |
| **EQE$_{max}$ [%]** | 12.6 | 14.2 | 12.8 | 12.0 | 13.4 |
| **L$_{max}$ ×10$^3$ [cd/m$^2$]** | 20.4 | 11.4 | 21.7 | 113.6 | 27.0 |
| **PE$_{max}$ [lm/W]** | 32.0 | 32.4 | 33.1 | 29.8 | 36.2 |
| **LE$_{max}$ [cd/A]** | 34.6 | 38.9 | 35.3 | 32.7 | 37.0 |
| **|J|$_{@5V}$ [mA/cm$^2$]** | 59.0 | 29.3 | 61.9 | 347.4 | 76.5 |

### 3.2. Reproducibility and consistency of the QLED EL performance

The operation of QLEDs in colling/heating cycles was repeated for different devices and the results were found to be consistent. Figure 5 displays the trend of EQE$_{max}$, L$_{max}$, V$_{on}$, and LE$_{max}$ at different temperatures for six QLEDs devices which were structurally identical. According to Figure 5a, EQE$_{max}$ of the devices first increased at -10 ℃, then it either remained unchanged, or reduced to near its initial value at RT2. Figure 5a, 5c, and 5d show that by increasing the temperature, EQE$_{max}$ and V$_{on}$, and LE$_{max}$ decreased, while the L$_{max}$ value increased (Figure 5b). The slight variations in the device performance parameters in the devices at each temperature arise from the thermal fluctuations in the Peltier plate during the measurements. Table 2 shows the average EQE$_{max}$, V$_{on}$, L$_{max}$ and LE$_{max}$ for six QLED devices with the standard deviations included at each temperature. These results show the consistency and reproducibility of our findings.

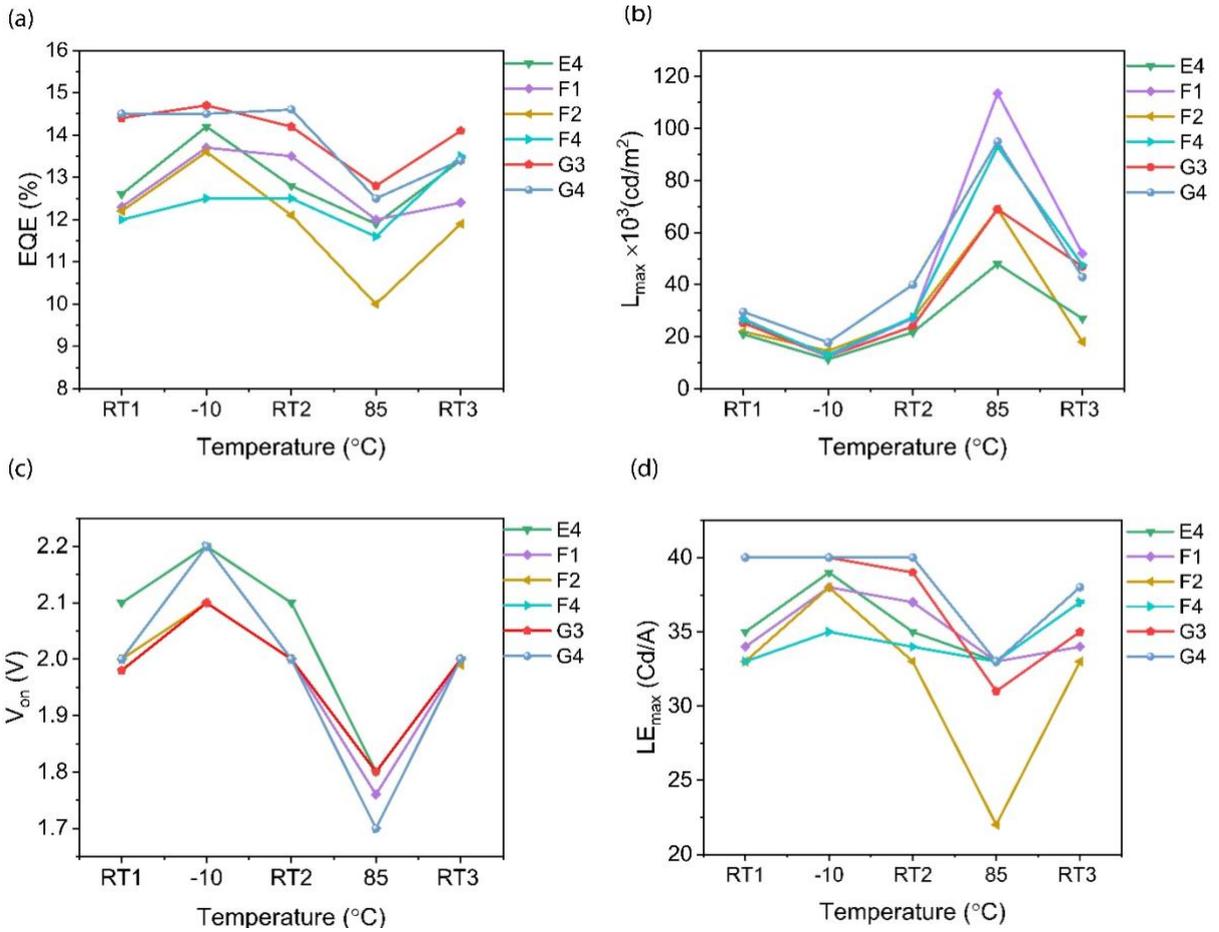

**Figure 5.** Effect of the temperature on (**a**) maximum external quantum efficiency (EQE$_{max}$), (**b**) maximum brightness, (**c**) Turn-on voltage, and (**d**) LE$_{max}$ of the amber QLED.

**Table 2.** Averaged efficiency parameters at various temperatures for six identical devices.

| T [°C] | EQE$_{max}$ [%] | L$_{max}$ ×10$^3$ [cd/m$^2$] @5v | V$_{on}$ [V] | LE$_{max}$ [cd/A] |
|---|---|---|---|---|
| **RT1** | 13.20±1.24 | 24.81±3.01 | 2.00±0.03 | 36.17±3.53 |
| **-10** | 14.00±1.15 | 13.11±2.81 | 2.20±0.05 | 38.17±2.96 |
| **RT2** | 13.40±1.16 | 28.20±5.20 | 2.00±0.04 | 36.50±3.20 |
| **80** | 12.00±0.90 | 79.44±25.41 | 1.80±0.05 | 31.33±3.90 |
| **RT3** | 13.10±0.88 | 41.63±11.55 | 2.00±0.03 | 35.17±2.17 |

Importantly, none of the devices failed under multiple thermal cycling. Moreover, the $EQE_{max}$ and $V_{on}$ in most of the devices were retrieved to near their initial values. An efficiency enhancement and drop was observed at -10 ℃ and a dropped 9% at 85 ℃ was while it recovered to its initial value after it cools down to RT3. In addition, $L_{max}$ was improved 68% at RT3 compared to $L_{max}$ at RT1. Although the $L_{max}$ dropped by 47% at -10 ℃, it recovered when the device reached the equilibrium with the ambient. $V_{on}$ showed a completely elastic behavior, such that it increased from 2 V to 2.2 V at -10 ℃ (reduction of carrier mobility) and recovered to 2 V at RT2. Followed by heating the device to 85 ℃, the turn-on voltage reduced to -1.8 V (enhanced carrier mobility at high T) and it recovered to 2 V at RT3. Finally, maximum current efficiency/luminance efficiency ($LE_{max}$) values are shown in the fourth column of Table 2. According to the $LE_{max} = L_{max}/J$, since the current density was minimum at -10 ℃, the current efficiency was the highest, 38 cd/A, when the temperature decreased. On the other hand, due to the maximum of current density at the elevated temperatures, the LE (L/J) is minimum, 31 cd/A, at 85 ℃. The current efficiency values almost recovered to their initial value of 36 cd/A (RT1), at each equilibration with the ambient temperature (RT2, RT3).

To better understand the effect of the temperature on the charge transport properties of the QLED devices, unipolar hole-only and electron-only devices (HOD and EOD) were fabricated, and their J-V curves were investigated in the cooling-heating cycles. Figure 6a and 6b demonstrates that the current density is significantly increased at 85 ℃ in both EOD and HOD. Rise in the temperature, facilitates carrier transport and increases the current density. Figures 6c-6d schematically show the HOD and EOD device structures, respectively.

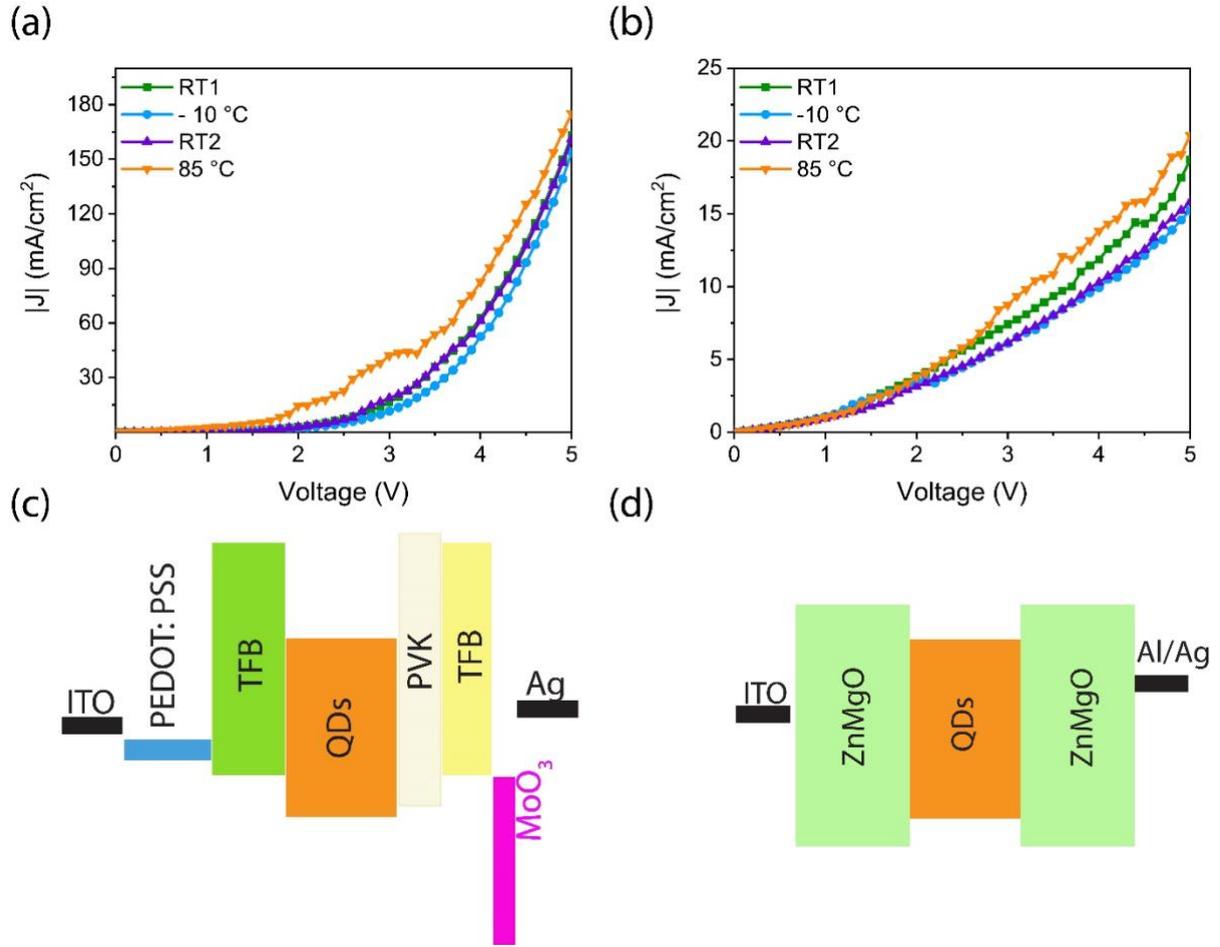

**Figure 6.** Effect of the temperature on the J-V curves of the **(a)** HOD, **(b)** EOD **(c)** the schematic layer stack of HOD and **(d)** EOD.

## 4. Conclusion

We have investigated the performance stability of our air-fabricated spin-coated amber-emitting QLED devices under thermal stress. We showed that our proposed QLED device performance is maintained at harsh weather conditions. The experiments were repeated for six different devices with consistent and reproducible results given by: The $EQE_{max}$, increased 6% at -10 ℃, and recovered at RT2. Although $EQE_{max}$ reduced 9% at 85 ℃, it recovered to near its initial value after equilibration to final room temperature (RT3). Moreover, at 85 ℃ the carrier transport performance was improved and $L_{max}$ enhanced by 220%, despite its 47% reduction at -10 ℃. In addition, $L_{max}$ increased by 68% at RT3 compared to $L_{max}$ at initial room temperature (RT1), which was due to a reduction of the density of the trap states after heating. The maximum current/luminance efficiency ($LE_{max}$) increased by 6% at -10 ℃, and reduced by 14% at 85 ℃, while $LE_{max}$ is recovered at each thermal equilibrium with room temperature (RT1, RT2, and RT3 values are almost equal). $V_{on}$ increased from 2 V to 2.2 V at -10 ℃ and recovered to 2 V at RT2. At 85 ℃, the turn-on voltage reduced to -1.8 V, then it recovered to 2 V at RT3. The improved carrier mobility at high temperatures was also confirmed by the results of the single-carrier EOD and HOD.

In the PL and the EL spectra, the peak position showed a blue shift from 601 nm at RT1 to 598 nm at -10 ℃, while it showed a red shift from 601 nm at RT2 to 606 nm at 85 ℃, and it recovered to 601 nm when it equilibrated to the RT3. The rise of the bandgap energy at low temperatures is the origin of the blue shift. The PL peak intensity was the highest at -10 ℃ and the lowest at 85 ℃, because the non-radiative recombination channels are reduced at low temperatures. However, the EL peak was intensified at high temperature, and diminished at the reduced temperature. The density of the trap states is reduced, and the number of the electrons escaped from the trapped states enhances. Therefore, the charge transport was improved, so the EL intensity, current density (J), and the brightness of the QLEDs at 85 ℃ increased, while the turn-on voltage decreased. The temperature-dependent behaviour of the unipolar HOD and EOD devices also justified our results. At lower temperatures, the reduction in the non-radiative recombination channels resulted in the improvement of the EQE at -10 ℃.

Due to their aerobic stability, our solution-processed QLEDs are cost effective and feasible to large-scale industrial fabrication, such as inkjet and roll-to-roll printing. Moreover, thermal stability of our QLEDs proves their suitability to operate at harsh weather conditions. The high and stable brightness levels of our QLEDs at any temperature make them suitable for applications such as indoor and outdoor lighting, as a cost-effective alternative for current OLEDs.

## 5. Experimental Section

**Materials.** Cadmium oxide (CdO, 99.99%, trace metals), trioctylphosphine (TOP, 90% technical grade), sulfur (S, 99.98%), 1-octadecene (ODE, 90%, technical grade), iso-octane (99.7%, HPLC grade), 1-octanethiol (>98.5%), trimethylammonium chloride (TMACl, >98%), potassium hydroxide (KOH, 99.99%), dimethyl sulfoxide (DMSO, >99.9%) magnesium acetate tetrahydrate (99%), zinc acetate dihydrate (>98%), lithium acetate (99.95%) and 1-butanol (anhydrous, 99.8%) were purchased from Sigma-Aldrich. Zinc acetate anhydrous (+99.9%) and ethyl acetate (>99.5%, ACS certified) were purchased from Thermoscientific. Selenium (Se, 99.999%, metals basis) and oleic acid (90%, technical grade) were purchased from Alfa-Aesar. Octane (+ 99%, extra pure) was purchased from Acros Organics. All the reagents were used as received. Poly(ethylene dioxythiophene): polystyrene sulfonate (PEDOT: PSS) was purchased from Ossila. Poly[(9,9-dioctylfluorenyl-2,7-diyl)-co-(4,4'-(N-(p-butylphenyl)) diphenylamine)] (TFB) was purchased from American Dye Source. Polyvinylpyrrolidone (PVP10) with an average molecular weight of 10000 was purchased from Sigma-Aldrich. Patterned ITO-glass with 15 Ω resistance and 25.4 mm×25.4 mm×0.7 mm was purchased from Luminescence Technology Corp.

**Synthesis of ZnCdSe core QDs.** CdO. zinc acetate. oleic acid and ODE were mixed with each other and degassed at 130°C. The temperature was raised to 300°C and selenium precursor in TOP was injected into the reaction flask. After core growth, the TOPSe precursor solution mixed with ODE was injected into the reaction flask (formation of ZnSe shell). Then, Se-S precursor was injected into the reaction. The sulfur solution (in TOP and ODE (1:3 vol%)), was added to octanethiol and injected into the flask. Then, the reaction was cooled down to 110°C. The QDs were washed with and reagent alcohol/acetonitrile, and redispersed in hexane. After the final step of the precipitation, the QDs were dispersed in octane.

**Synthesis of Li-Mg-doped ZnO NPs.** TMAH was synthesized by dissolving 2.2 g of TMACl in 14 ml of reagent alcohol, and 1.1 g of KOH in 16 ml of reagent alcohol, separately. After the complete dissolving, the KOH and TMACl solutions were mixed and centrifuged at 5000 rpm for 3 min. The resulting TMAH was filtered using 0.22 μm PTFE filters, while the KCl solid part was separated. Next, 10%Li:10%Mg:ZnO (LMZO) was synthesized by solution-precipitation method with some modifications [48]. 8 mmol of zinc acetate dihydrate, was mixed with 1 mmol of magnesium acetate dihydrate, and 1 mmol of lithium acetate hydrate, and dissolved in 25 ml of dimethyl sulfoxide (DMSO). Then the temperature of the solution was reduced to below 2 °C and 21 ml of TMAH was added dropwise, and the solution was stirred for 2 h at T < 2 °C. The nanoparticles were then washed twice with ethyl acetate and redispersed in butanol.

**QLED device fabrication.** All the device spin-coating processes were done in-air, at room temperature with a moisture level of 11-15% in winter, and 25-35% in summer. The ITO glass substrates were sonicated in a bath of detergent and DI water, acetone, and IPA. The substrate surfaces were then treated with a UV-Ozone lamp for 15 minutes. The PEDOT:PSS HIL was spin-coated at 5000 rpm for 40 sec, and annealed at 130 °C for 20 minutes. 8 mg/ml TFB in toluene solution was spin-coated at 3000 rpm for 40 sec, and annealed at 110 °C for 20 min. ZnCdSe/ZnSe/ZnSeS/ZnS QDs were spin-coated at 3000 rpm for 35 sec, and annealed at 80 °C for 15 min. 5wt% PVP was dissolved in 30 mg/ml LMZO and spin-coated at 3000 rpm for 35 sec, then annealed at 80 °C for 20 min. Finally, 50 nm Al and 120 nm Ag was deposited as the anode by thermal evaporation. The devices were then encapsulated with UV-curable epoxy.

**EOD fabrication.** After cleaning the ITO substrates, LMZO doped with PVP was spin-coated at 3000 rpm for 35 sec, then annealed at 80 °C for 20 min. QDs were spin-coated at 3000 rpm for 35 sec, and annealed at 80 °C for 15 min. The second layer of the doped ZnO NPs was spin-coated on top of the QDs at 3000 rpm, for 35 sec, and annealed at 80 °C for 20 minutes. Al and Ag were deposited by thermal evaporation. The devices were then encapsulated with UV-curable epoxy.

**HOD fabrication.** On the clean and UV-Ozone-treated ITO substrates, PEDOT:PSS HIL was spin-coated at 5000 rpm for 40 sec, and annealed at 130 °C. 8 mg/ml TFB in toluene solution was spin-coated at 3000 rpm for 40 sec, and annealed at 110 °C for 20 min. QDs were spin-coated at 3000 rpm for 35 sec, and annealed at 80 °C for 15 min. 5 mg/ml of PVK in 1,4-dioxane was spin-coated at 4000 rpm for 40 sec, and annealed at 100 °C for 20 minutes. 5 mg/ml TFB in p-xylene was spin-coated at 4000 rpm for 40 sec and annealed at 100 °C for 20 minutes. The samples were then transferred to the thermal evaporation device, while 8 nm of $MoO_3$, and 200 nm of Ag were deposited by thermal evaporation. The devices were then encapsulated with UV-curable epoxy.

**Characterization**

**Materials.** The PL spectrum of the QDs was characterized using StellarNet Miniature UV-Vis spectrometer. The absorption spectrum of the QDs and LMZO NPs were measured using Cary 60 UV-Vis spectrometer from Agilent Technologies. The QLED layer stack structure and the thickness of each layer was characterized by TEM-cross sectional image. TEM and S/TEM/EDX analysis was performed on a JEOL ARM 200cf microscope, which is equipped with a cold-field emitter and probe Cs corrector and operates at 200 kV acceleration voltage.

**Devices.** QLED EL parameters were measured using StellarNet Miniature UV-Vis spectrometer. and the QLED performance was characterized by a Si-photodetector connected to a Keithly 2612B source meter, while the devices were pre-biased at 7 V before the testing, with a DC power supply. The device was run and the data was collected using a Python code, by which the voltage was swept from -2 V to 12 V and $EQE_{max}$, $L_{max}$, J, V, $PE_{max}$ and $LE_{max}$ were calculated and collected using the QDs EL parameters.

**Temperature-dependent PL, EL, QLED, EOD, and HOD testing.** QLEDs were put on TEC1-12715 0-15V heatsink thermoelectric Peltier cooler with a piece of a 0.5 mm thermally conductive pad, from Thermal Right Co. The positive and negative poles of the Peltier plate were connected to the corresponding poles of the DC supply. By applying voltage, the Peltier heated up and the QLED temperature, which was in contact with the thermal pad, increased. The temperature of the device surface was carefully monitored with a FLIR E8-XT infrared thermal camera, till it reached 85 °C. The device was then operated, and the performance parameters were collected with a Si photodiode. For low-temperature testing, the connection of the positive and negative poles was switched, such that the surface of the Peltier cooled down to -10 °C. In the experiments, the low-temperature limit (-10°C) was imposed by the limitations of the cooling capacity of the Peltier plate. Moreover, the upper temperature (85°C) was limited by the thermal degradation of the UV-curable epoxy which was in direct contact with the cathode and the glass coverslip, causing a color-yellowing in the epoxy and observable damage to the devices at temperatures higher than 85°C. Of course, adopting a non-contact encapsulation solution with the epoxy resin could make it possible to perform the measurement at higher temperatures.

## Acknowledgments

M.P. acknowledges MITACS accelerate program (project no. IT17040) for financial support. S.K.-C. acknowledges support from the Canada Research Chairs Program. The authors sincerely thank Peng Li, Xuehai Tan, and Shihong Xu from NanoFab (University of Alberta) for cross-sectional TEM measurement.

## Conflict of Interest

The authors declare no conflict of interest.